\documentclass[12pt,preprint]{emulateapj}
\usepackage{graphicx}

\shorttitle{Inclination Dependence of Phase Signatures}
\shortauthors{Stephen R. Kane \& Dawn M. Gelino}
\slugcomment{Submitted for publication in the Astrophysical Journal}

\begin{document}

\title{On the Inclination Dependence of Exoplanet Phase Signatures}
\author{Stephen R. Kane, Dawn M. Gelino}
\affil{NASA Exoplanet Science Institute, Caltech, MS 100-22, 770
  South Wilson Avenue, Pasadena, CA 91125, USA}
\email{skane@ipac.caltech.edu}


\begin{abstract}

Improved photometric sensitivity from space-based telescopes have
enabled the detection of phase variations for a small sample of hot
Jupiters. However, exoplanets in highly eccentric orbits present
unique opportunities to study the effects of drastically changing
incident flux on the upper atmospheres of giant planets. Here we
expand upon previous studies of phase functions for these planets at
optical wavelengths by investigating the effects of orbital
inclination on the flux ratio as it interacts with the other effects
induced by orbital eccentricity. We determine optimal orbital
inclinations for maximum flux ratios and combine these calculations
with those of projected separation for application to coronagraphic
observations. These are applied to several of the known exoplanets
which may serve as potential targets in current and future coronagraph
experiments.

\end{abstract}



\section{Introduction}
\label{introduction}

The changing phases of an exoplanet as it orbits the host star has
long been considered as a means for their detection and
characterization \citep{cha99,lei03}. The relation between giant
planet atmospheres and phase curves have been described in detail by
\citet{sud05} and \citet{kan10}, hereafter KG10. However, the
relatively small flux ratio of the planet to the host star has
presented a major hinderance to the realization of such detections. A
new era of optical and Infra-Red (IR) telescopes are enabling phase
detections in a manner which was previously inaccessible from the
ground.

Attempts to detect phase signatures have primarily been for transiting
planets. Examples of observed phase variations in the IR include
HD~189733b \citep{knu09a} and HD~149026b \citep{knu09b}. Examples in
the optical include Kepler observations of HAT-P-7b \citep{wel10} and
phase variations detected in the light curve of CoRoT-1b
\citep{sne09}. Phase variation detection of non-transiting planets
have been restricted to hot Jupiters, including $\upsilon$~And~b
\citep{har06} and HD~179949b \citep{cow07}. There has also been recent
observational evidence for the detection of phase variations by the
non-transiting planet HD~46375b by \citet{gau10}. Planets in eccentric
orbits, such as HD~17156b \citep{bar07} and HD~80606b \citep{lau09},
will produce relatively high phase amplitudes during a brief interval.

The orbital inclination is usually preferred to be edge-on for optimal
detection of phase variations since this ensures that full phase will
occur along the line-of-sight. However, this is not necessarily the
case depending upon the eccentricity and periastron argument of the
orbit, and indeed there are cases where the maximum flux ratio occurs
when the orbit is face-on. Even though the radial velocity technique
is biased towards the detection of edge-on orbits, since these have a
larger radial velocity semi-amplitude, certainly this is not always
the case. Astrometric studies such as that performed by \citet{han01}
have shown that there are a variety of orbital inclinations as
expected, and so it is prudent to consider this for attempted phase
detections.

Here we present a thorough exploration of orbital parameter-space and
the impact upon flux ratios for eccentric planets. The main causes of
degeneracy are the periastron argument and inclination for a given
eccentricity. We further calculate projected separations at apastron
as a function of inclination and determine their correspondance with
maximum flux ratio locations. With all orbital elements considered,
one may then use the phase locations of maximum flux ratio and maximum
projected separation to quantify the suitability for follow-up
observations.


\section{Flux Ratio Components}

In this section, we outline the theoretical framework which will be
used through the remainder of the paper. This formalism has been used
extensively by \citet{col02} and more recently by \citet{rod10}. We
refer the reader to KG10 for a more detailed description of
the particular formalism used here.

The flux ratio of a planet with radius $R_p$
to the host star is defined as
\begin{equation}
  \epsilon(\alpha,\lambda) \equiv
  \frac{f_p(\alpha,\lambda)}{f_\star(\lambda)}
  = A_g(\lambda) g(\alpha,\lambda) \frac{R_p^2}{r^2}
  \label{fluxratio}
\end{equation}
where the flux is measured at wavelength $\lambda$.  This flux ratio
consists of three major components; the geometric albedo
$A_g(\lambda)$, the phase function $g(\alpha,\lambda)$, and the
inverse-square relation to the star--planet separation $r$. Each of
these are briefly described below in the context of eccentric orbits.
Note that the observed flux ratio from an exoplanet is wavelength
dependent in that the atmospheric composition drives the scattering
properties and thus the shape of the geometric albedo and phase
function. As was the case with KG10, we confine our study to optical
wavelengths centered on 550~nm.

Atmospheric models have shown that there is a dependence of the
geometric albedo of giant planets on the semi-major axis of the orbit
\citep{sud00,sud05,cah10}. To account for this, we use the analytic
function described by KG10 which was in turn derived from the models
of \citet{sud05}. This function results in a time-dependence
of the albedo as the strong irradiation of the atmospheres of giant
planets removes reflective condensates from the upper atmospheres
during periastron passage. This function can have a dramatic effect in
dampening the flux ratio at small star--planet separations.

The phase angle $\alpha$ is defined to be zero when the planet is at
superior conjunction and is described by
\begin{equation}
  \cos \alpha = \sin (\omega + f)
  \label{phaseangle}
\end{equation}
where $\omega$ is the argument of periastron and $f$ is the true
anomaly.  Thus, in terms of orbital parameters, minimum phase occurs
when $\omega + f = 90\degr$ and maximum phase occurs when $\omega + f
= 270\degr$.

The exact nature of a planetary phase function depends upon the
assumptions regarding the scattering properties of the
atmosphere. Rather than assuming isotropic scattering (Lambert
sphere), we adopt the empirically derived phase function of
\citet{hil92} which is based upon observations of Jupiter and Venus
and incorporates substantially more back-scattering due to
cloud-covering. This approach contains a correction to the planetary
visual magnitude of the form
\begin{equation}
  \Delta m (\alpha) = 0.09 (\alpha/100\degr) + 2.39
  (\alpha/100\degr)^2 -0.65 (\alpha/100\degr)^3
  \label{magcorr}
\end{equation}
leading to a phase function given by
\begin{equation}
  g(\alpha) = 10^{-0.4 \Delta m (\alpha)}
  \label{phase}
\end{equation}
which is used throughout the remainder of this paper.

As shown by \citet{sud05} and KG10, the maximum flux ratio does not
necessarily occur at zero phase angle for a non-circular orbit. This
is because the star--planet separation $r$ is constantly changing and
is given by
\begin{equation}
  r = \frac{a (1 - e^2)}{1 + e \cos f}
  \label{separation}
\end{equation}
where $a$ is the semi-major axis and $e$ is the orbital eccentricity.
The $R_p^2/r^2$ component of Equation \ref{fluxratio} becomes dominant
for highly eccentric orbits which is especially important when the
orbital inclination is considered.


\section{Orbital Inclination}
\label{inclination}

To add the effect of inclination angle to the phase function, the
phase angle (Equation \ref{phaseangle}) is modified as follows:
\begin{equation}
  \cos \alpha = \sin (\omega + f) \sin i
  \label{phaseinc}
\end{equation}
At 1st and 3rd quarter ($\alpha = 90\degr$ and $\alpha = 270\degr$),
the flux ratio is completely independent of inclination
angle. However, at all other phase angles there arises a complex
pattern of flux ratios from the inclination and argument of
periastron. Here we describe these dependencies.


\subsection{Edge-on/Face-on Orbits}

The two cases worth considering first are the extremes of edge-on ($i
= 90\degr$) and face-on ($i = 0\degr$) orbits. As noted earlier, the
radial velocity method is biased towards the detection of edge-on
orbits due to the increase in the semi-amplitude of the signal. In
addition, most of the planets monitored for phase signatures are known
to transit. Thus the case of edge-on orbits is currently the dominant
form of investigated systems. This guarantees the observability of
both zero phase and full phase, the contribution of which to the flux
ratio depends upon the star--planet separation at $\alpha = 0\degr$.

The case of face-on orbits means that the phase function becomes
completely flat since only half phase will be visible at any one
time. Thus the flux ratio is completely determined by the eccentricity
of the orbit which drives both the star--planet separation and the
changing albedo of the upper atmosphere.


\subsection{Generalized Orientation}

\begin{figure*}
  \begin{center}
    \begin{tabular}{cccc}
      \includegraphics[angle=270,width=4.1cm]{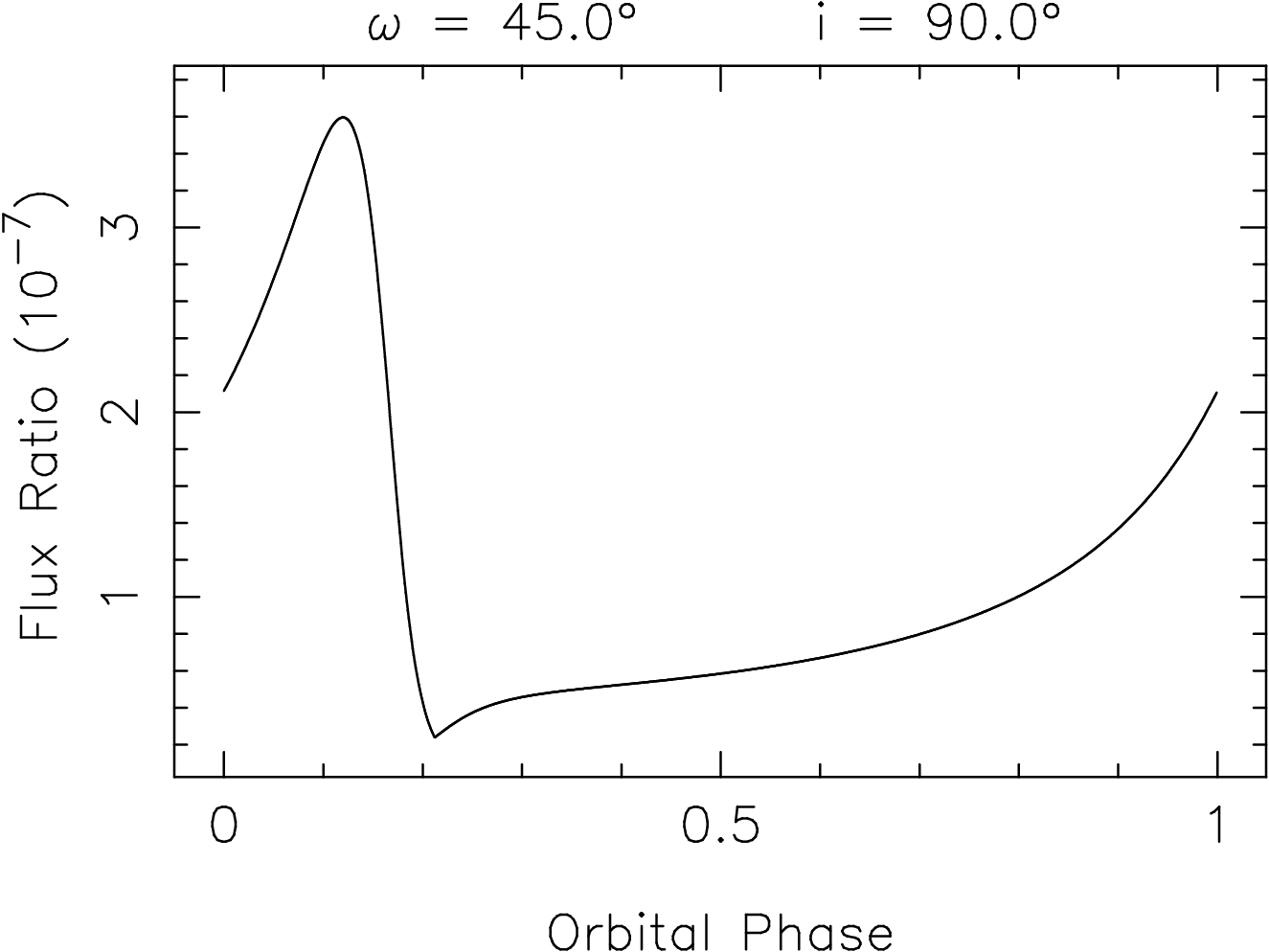} &
      \includegraphics[angle=270,width=4.1cm]{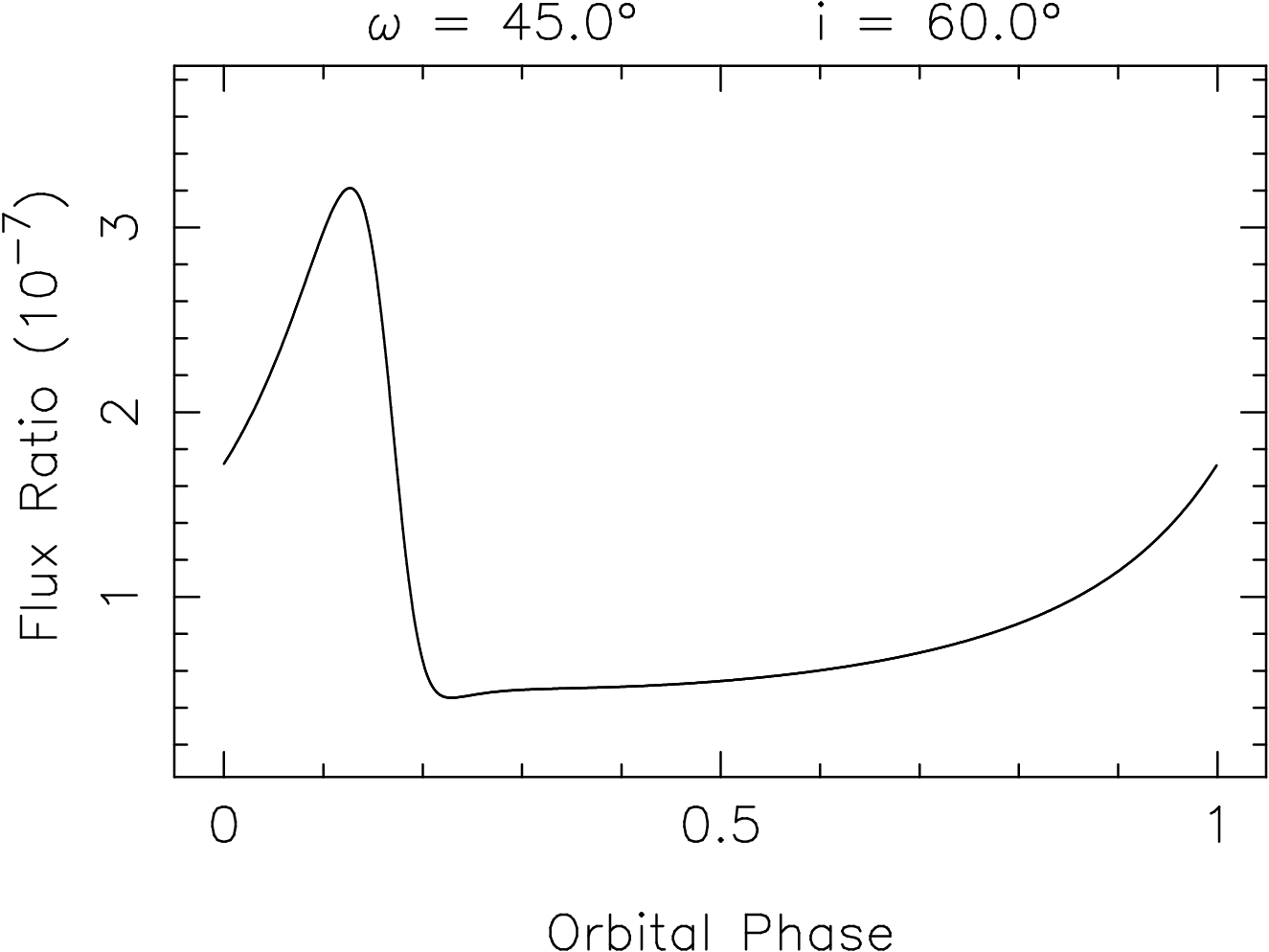} &
      \includegraphics[angle=270,width=4.1cm]{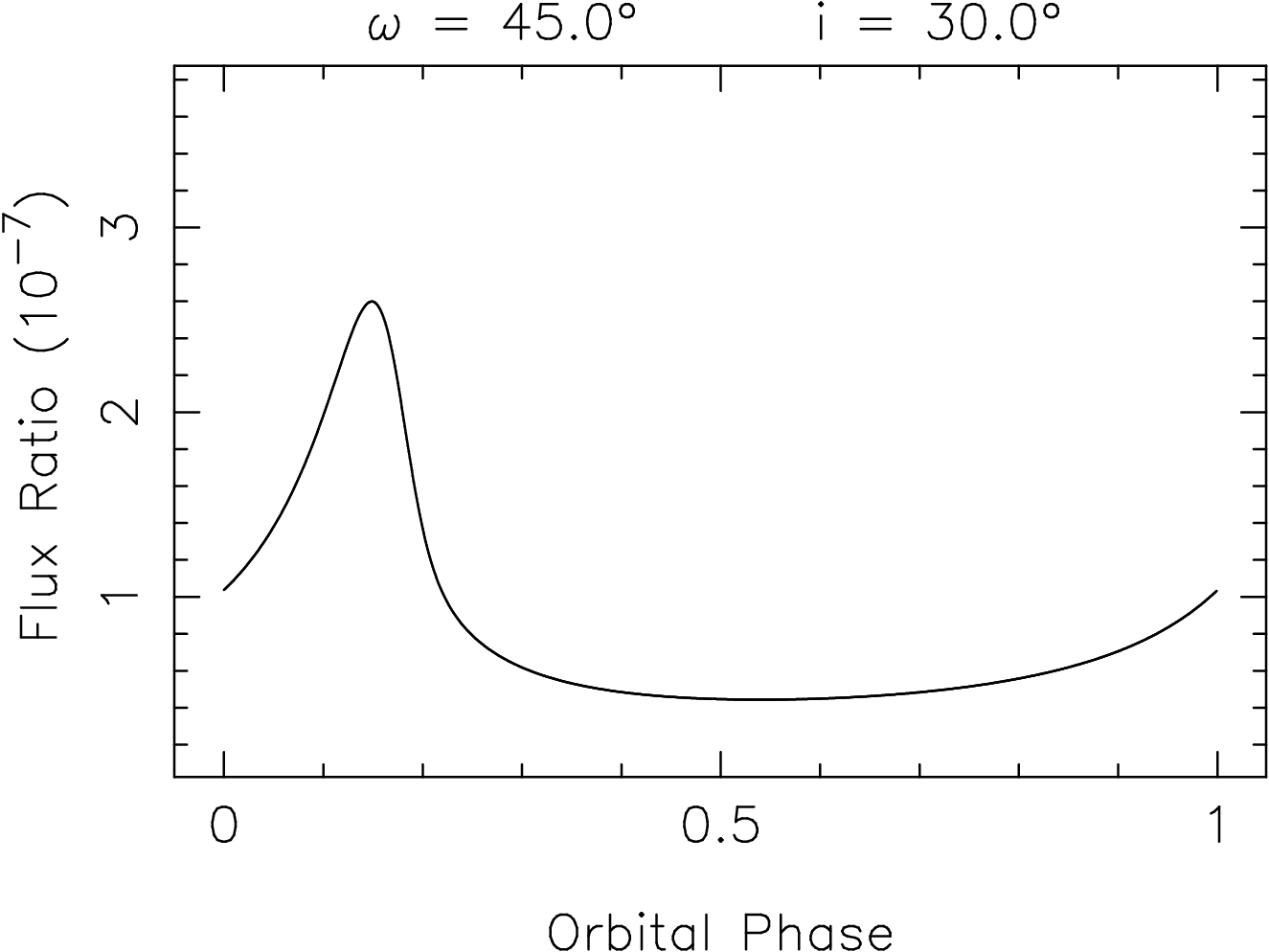} &
      \includegraphics[angle=270,width=4.1cm]{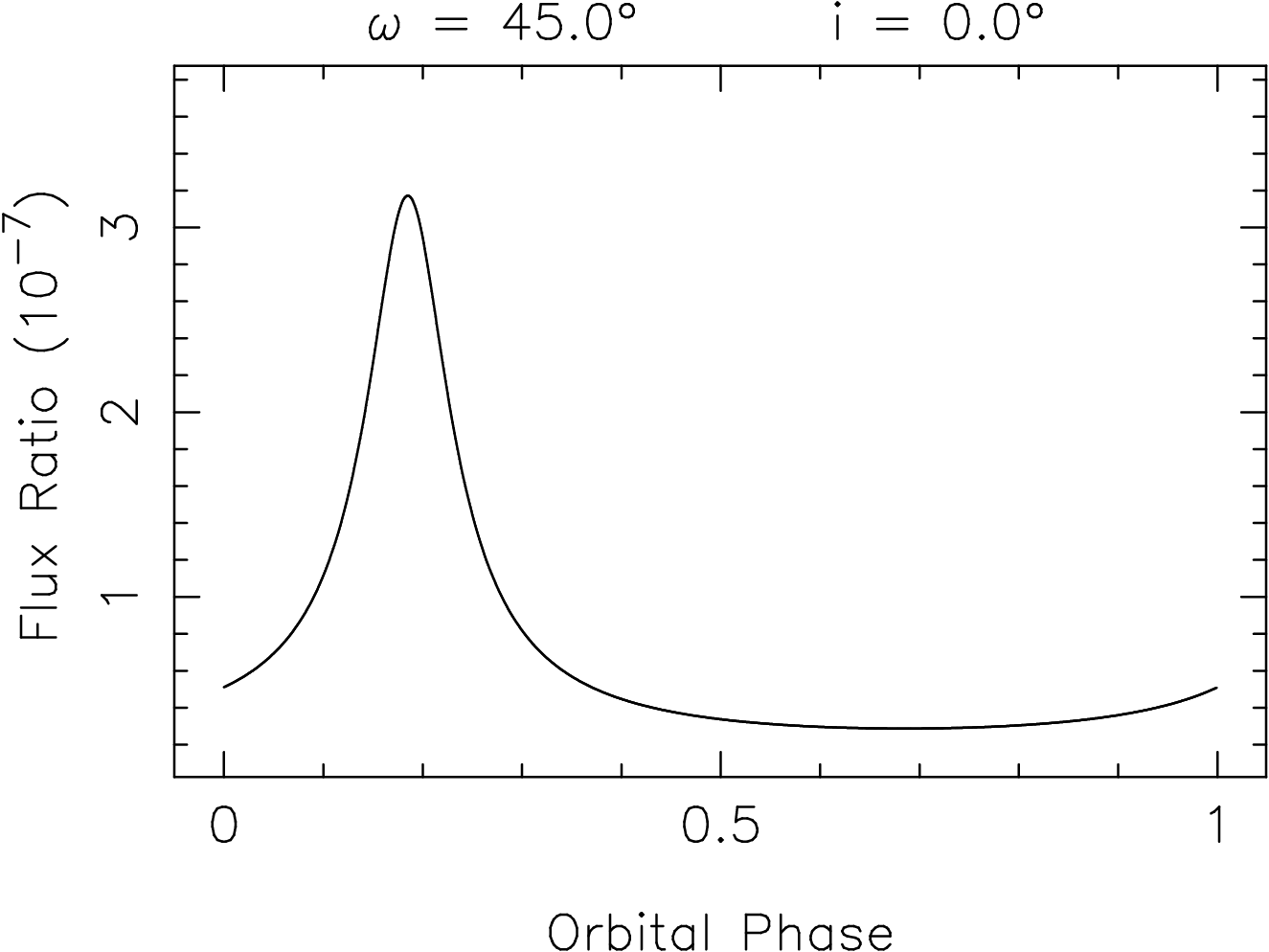} \\
      \includegraphics[angle=270,width=4.1cm]{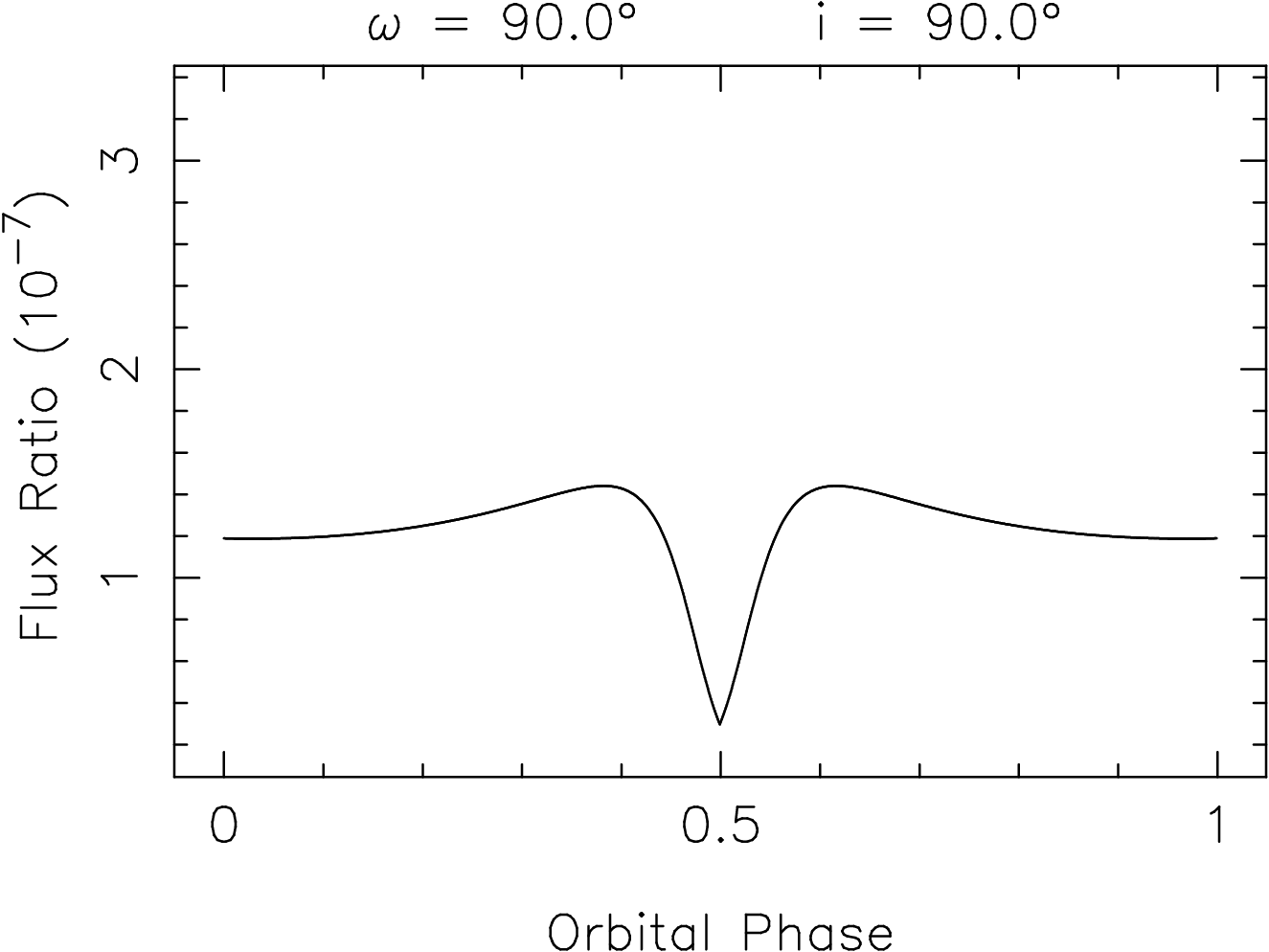} &
      \includegraphics[angle=270,width=4.1cm]{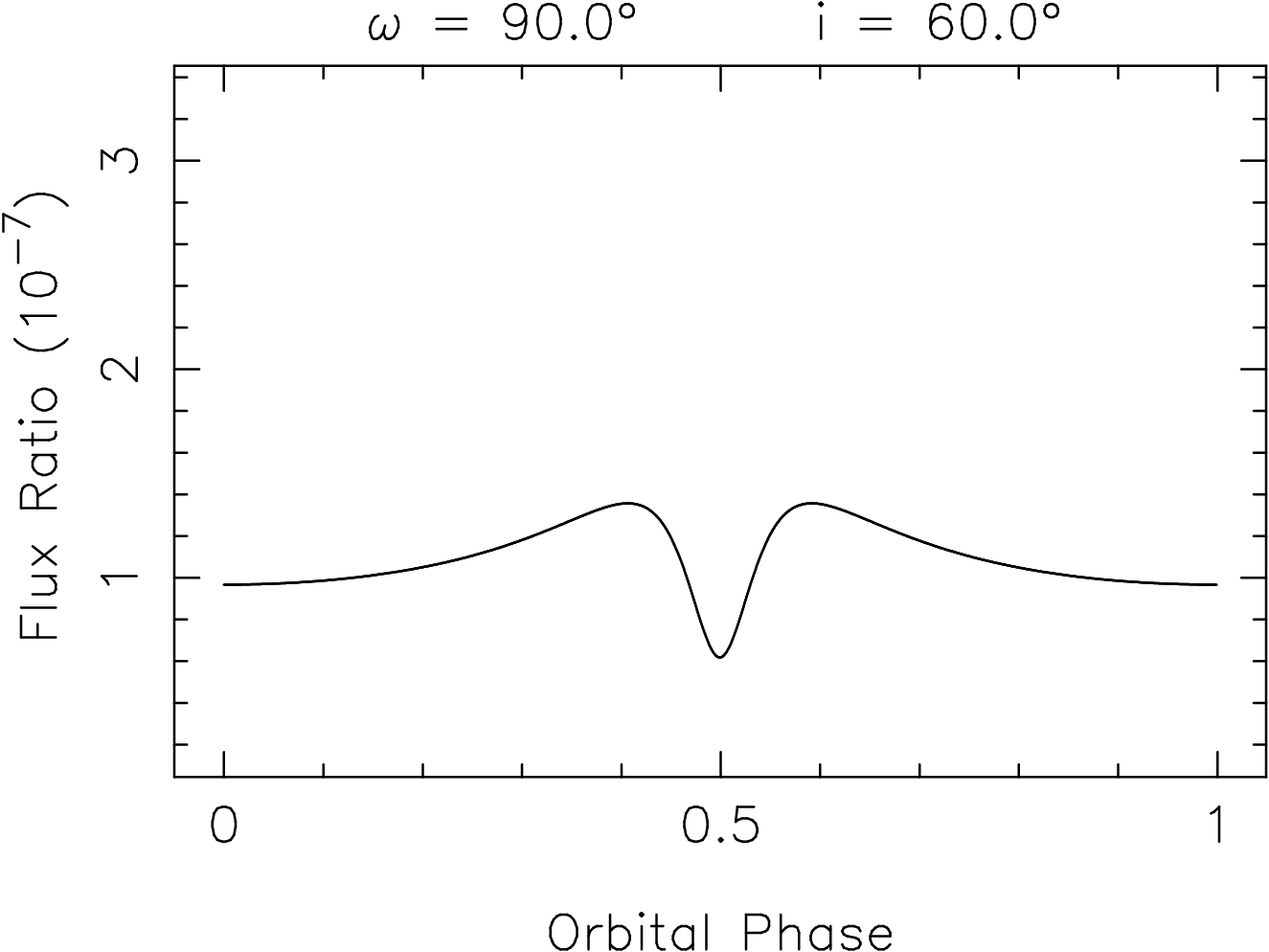} &
      \includegraphics[angle=270,width=4.1cm]{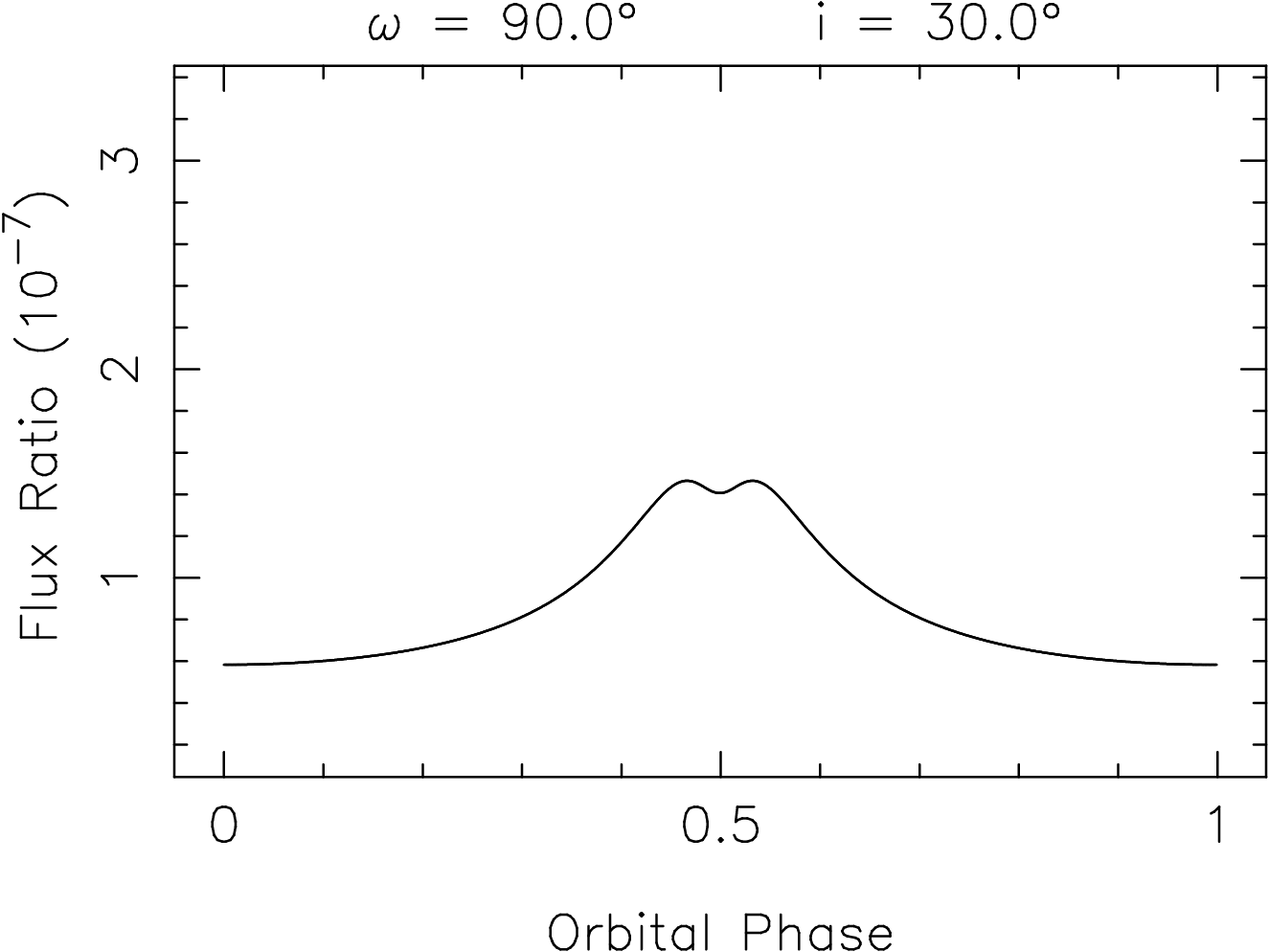} &
      \includegraphics[angle=270,width=4.1cm]{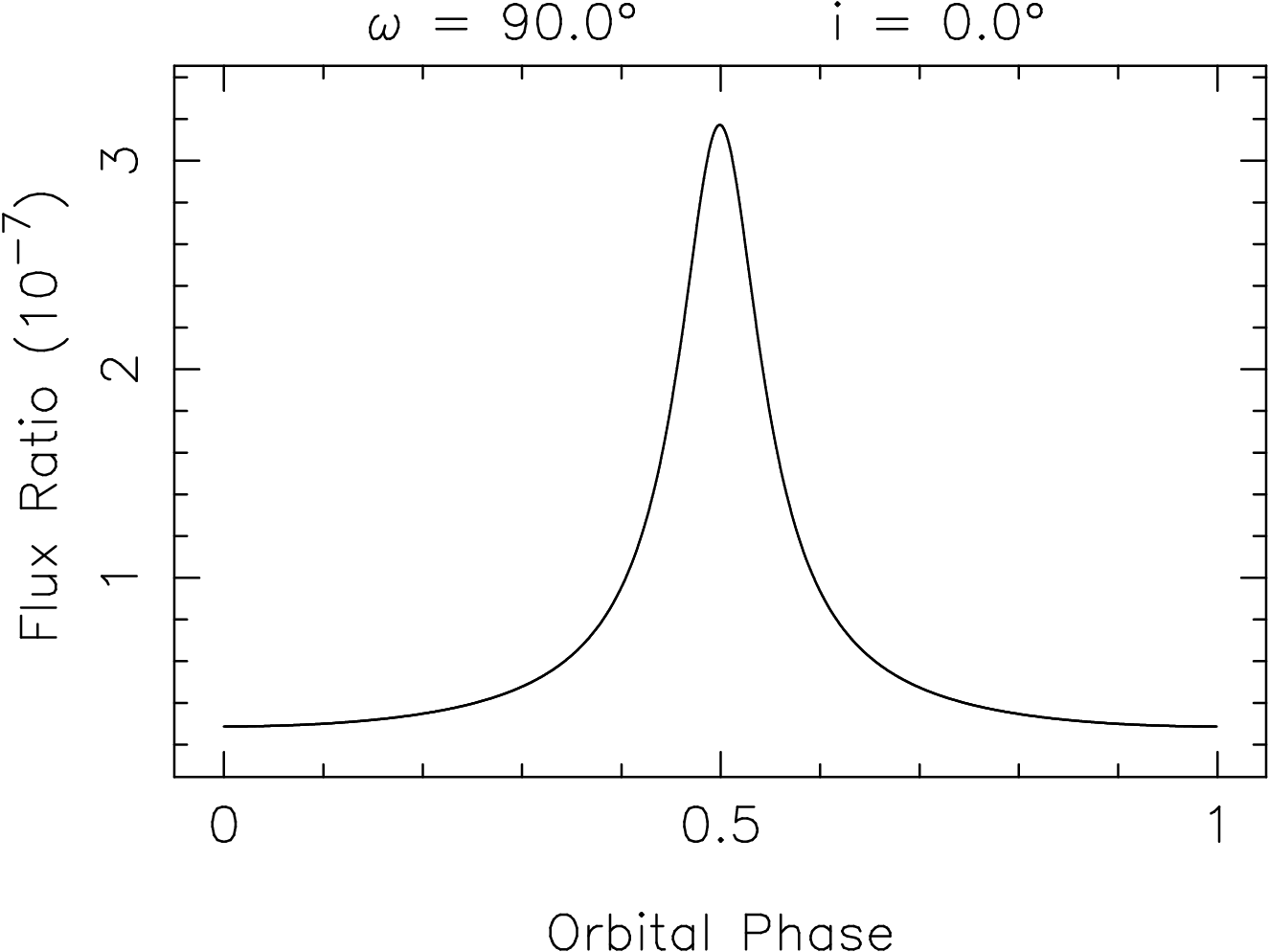} \\
      \includegraphics[angle=270,width=4.1cm]{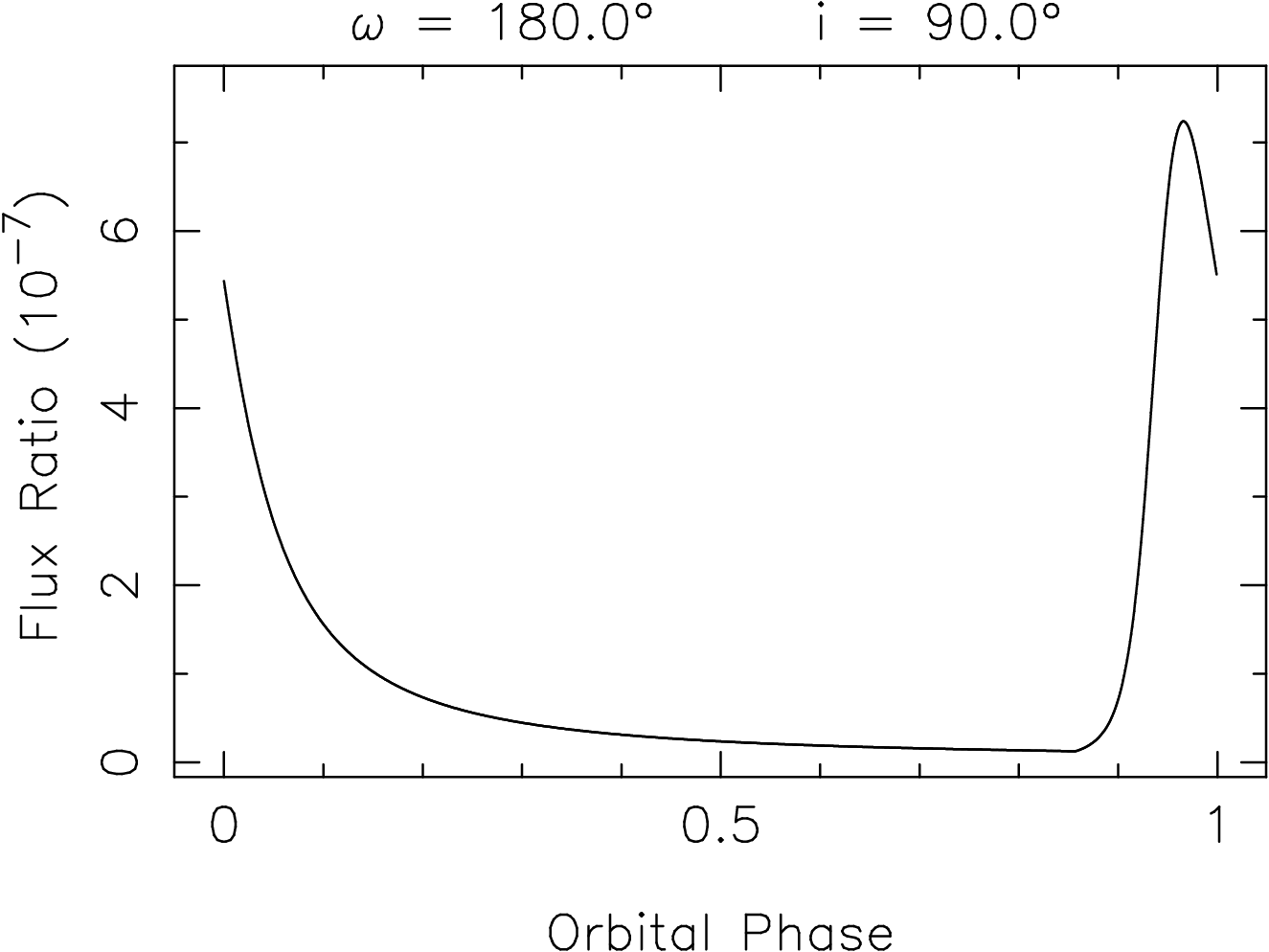} &
      \includegraphics[angle=270,width=4.1cm]{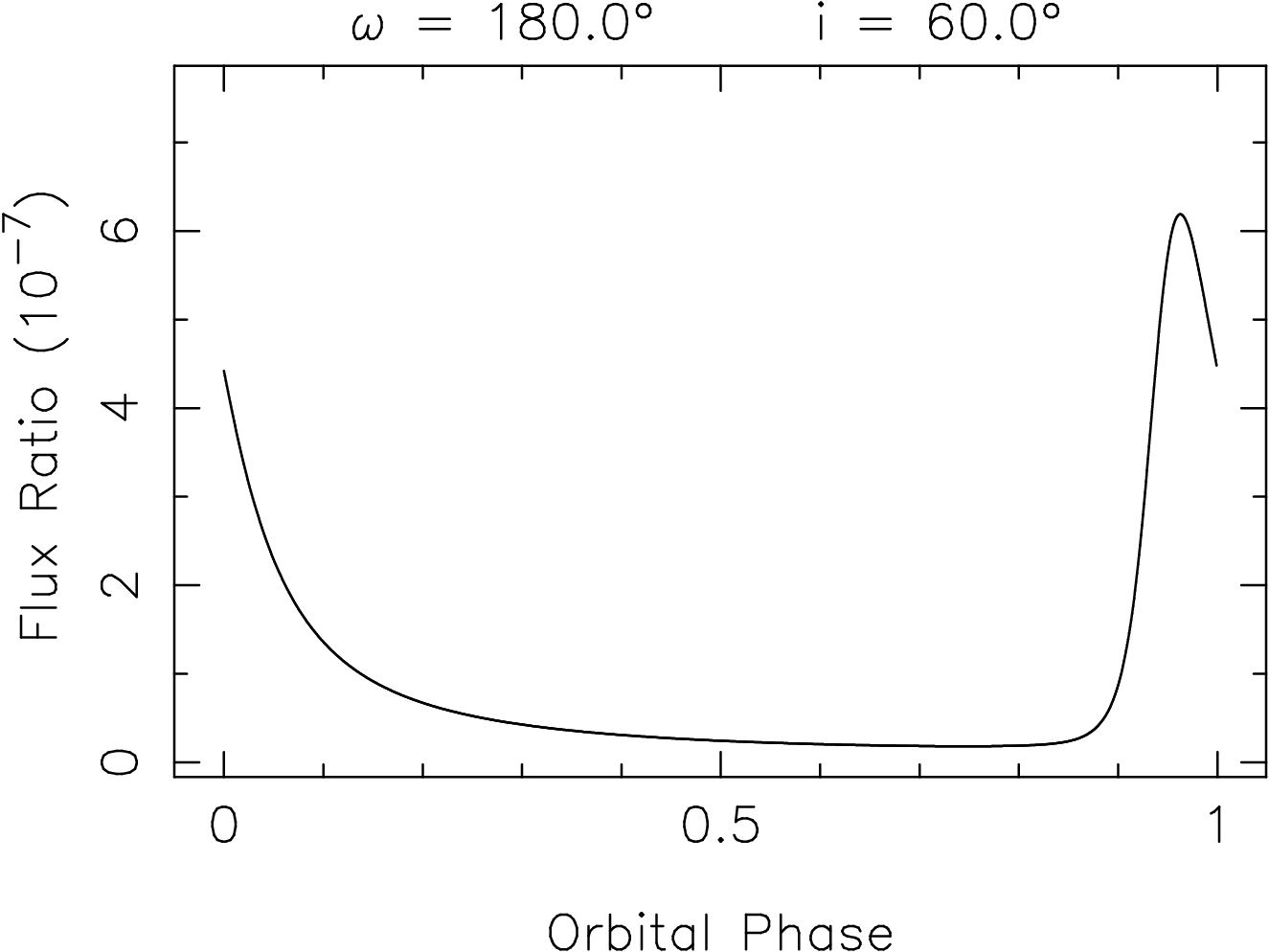} &
      \includegraphics[angle=270,width=4.1cm]{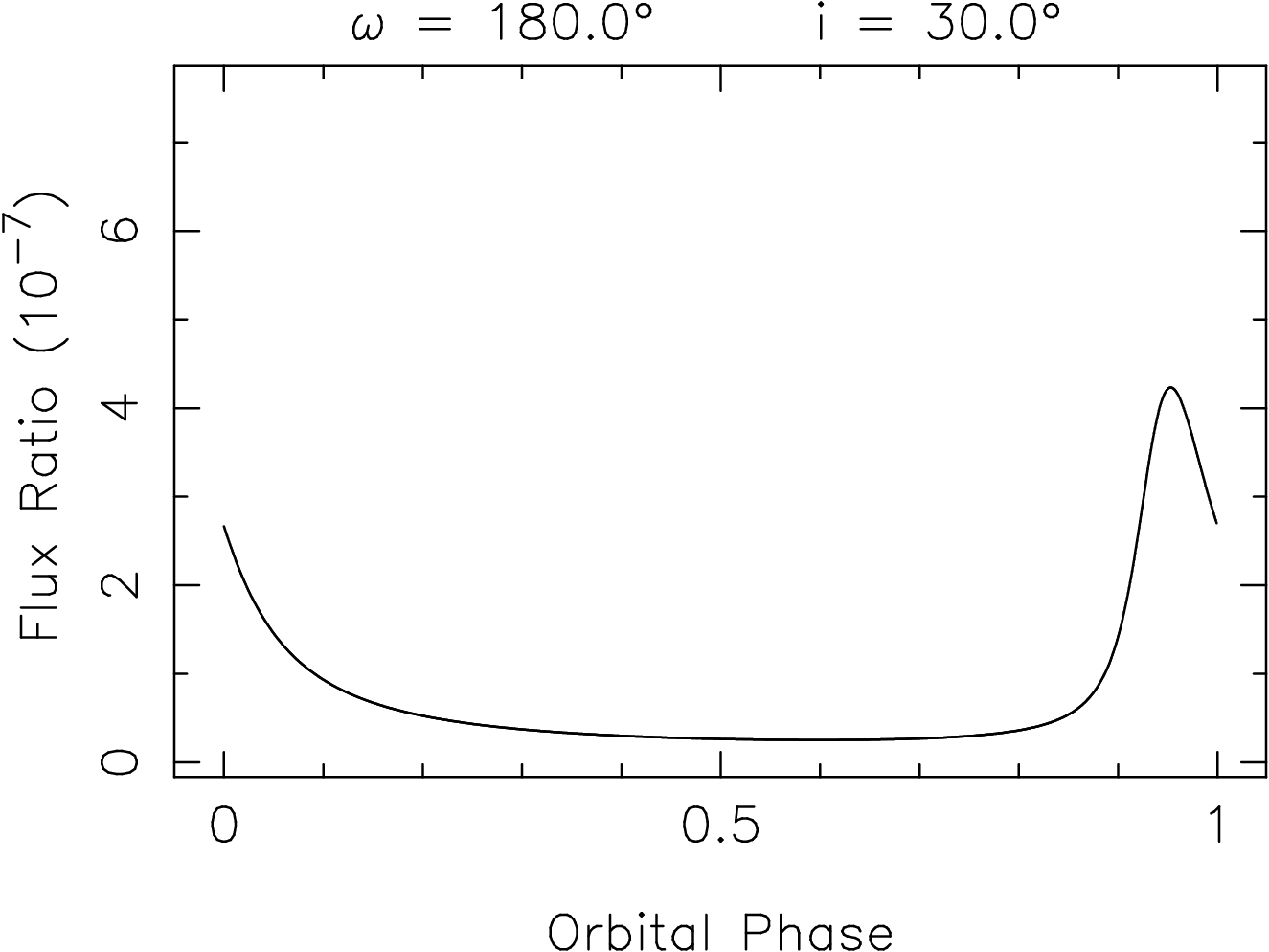} &
      \includegraphics[angle=270,width=4.1cm]{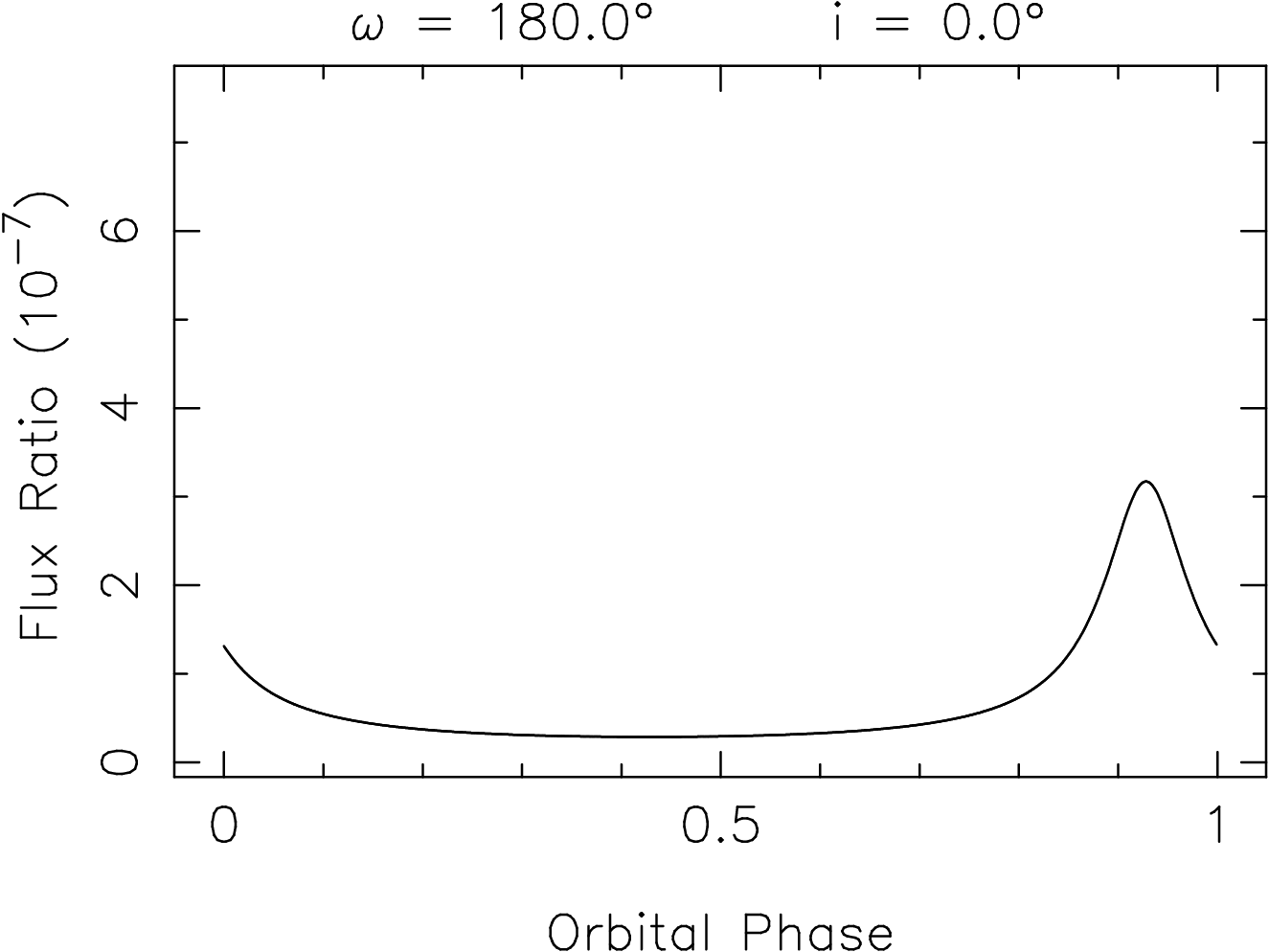} \\
      \includegraphics[angle=270,width=4.1cm]{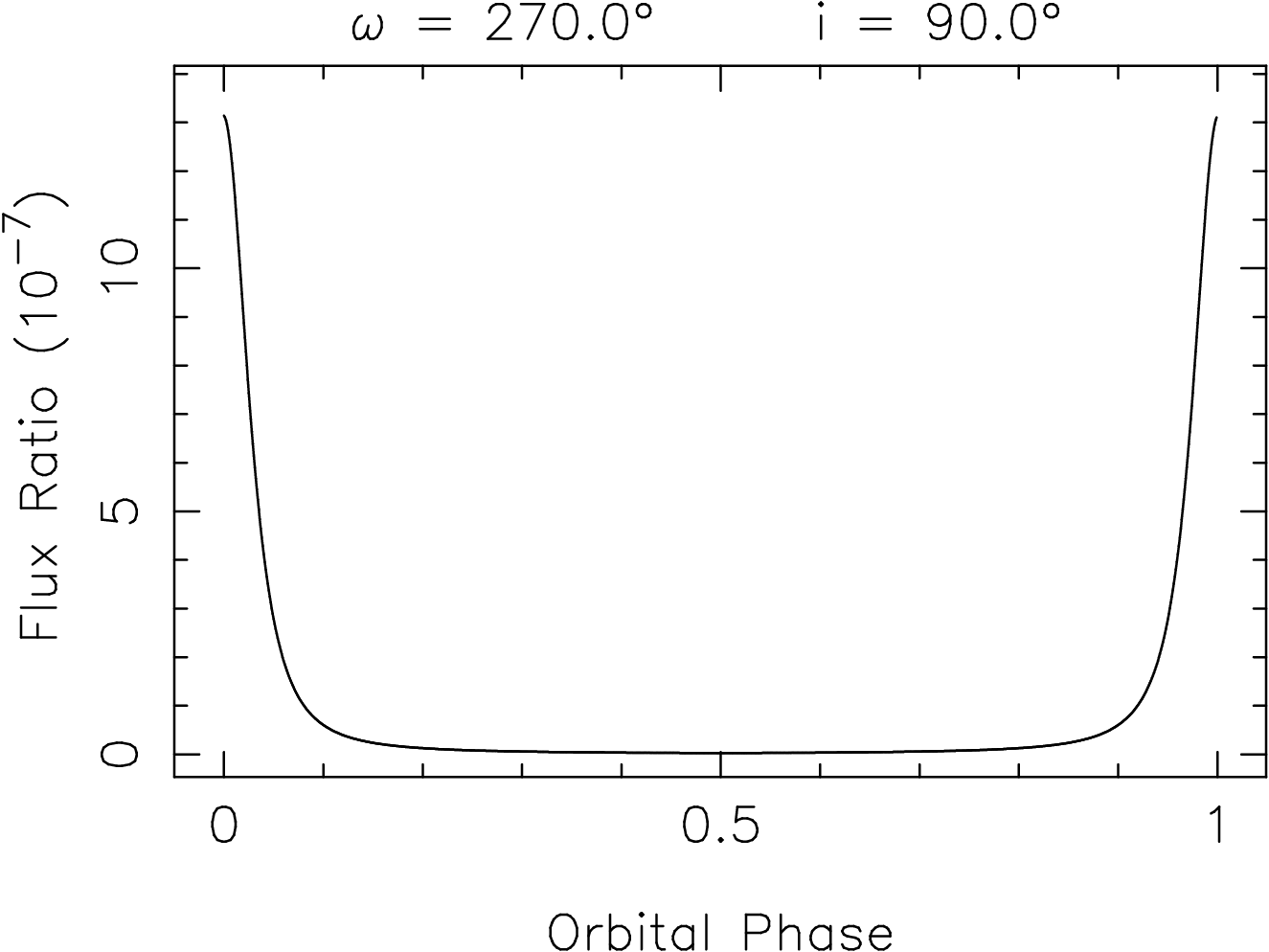} &
      \includegraphics[angle=270,width=4.1cm]{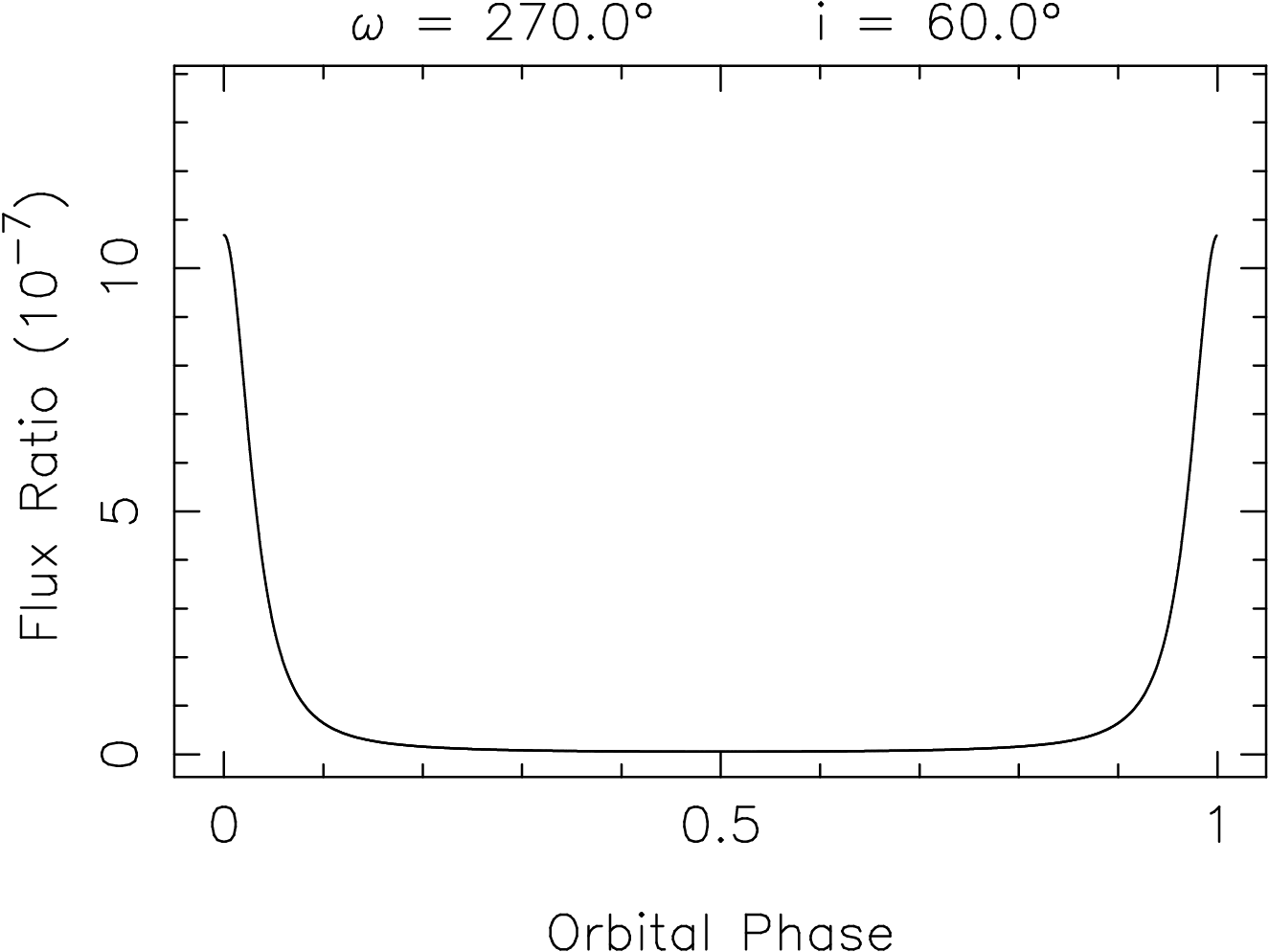} &
      \includegraphics[angle=270,width=4.1cm]{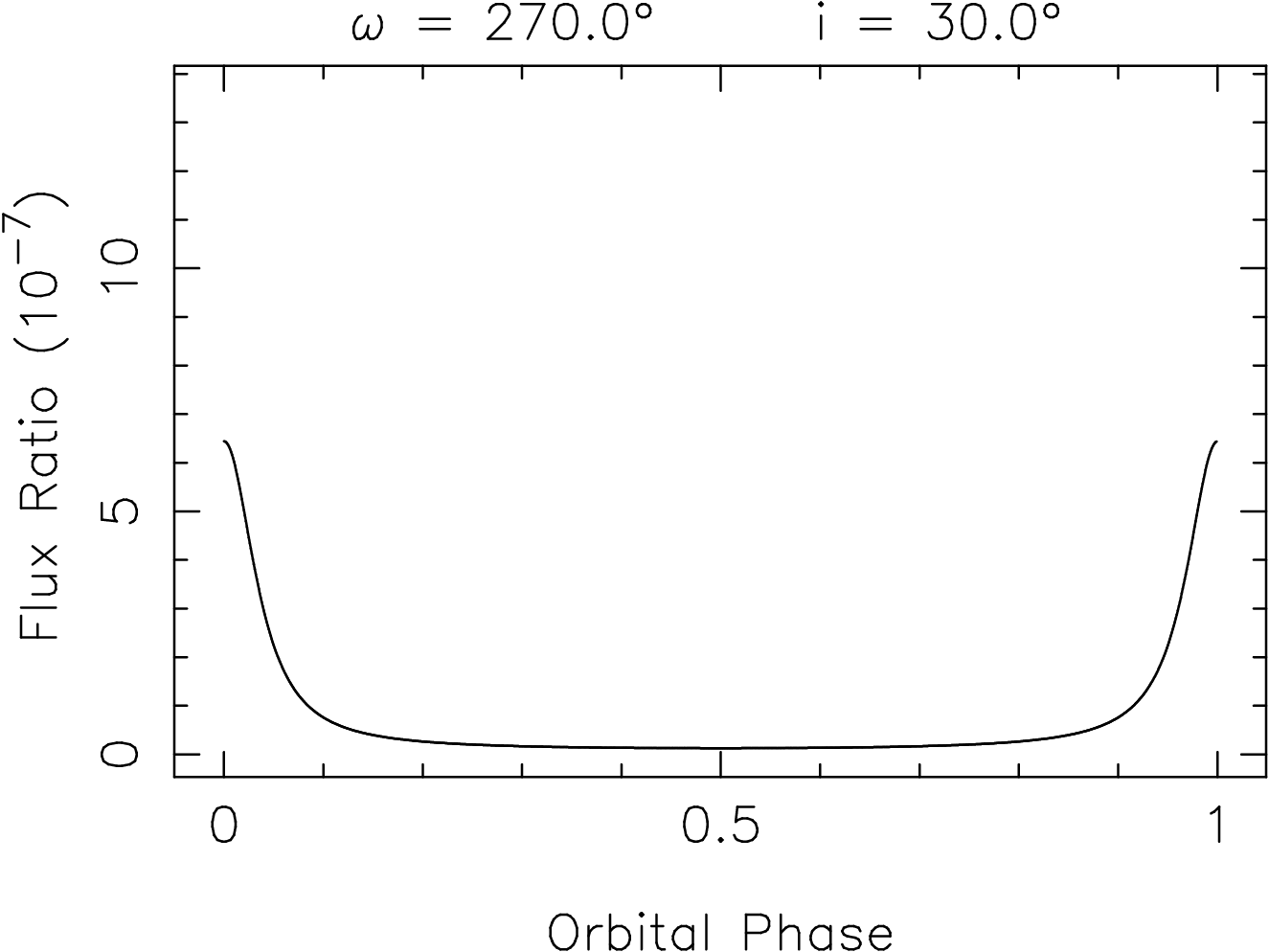} &
      \includegraphics[angle=270,width=4.1cm]{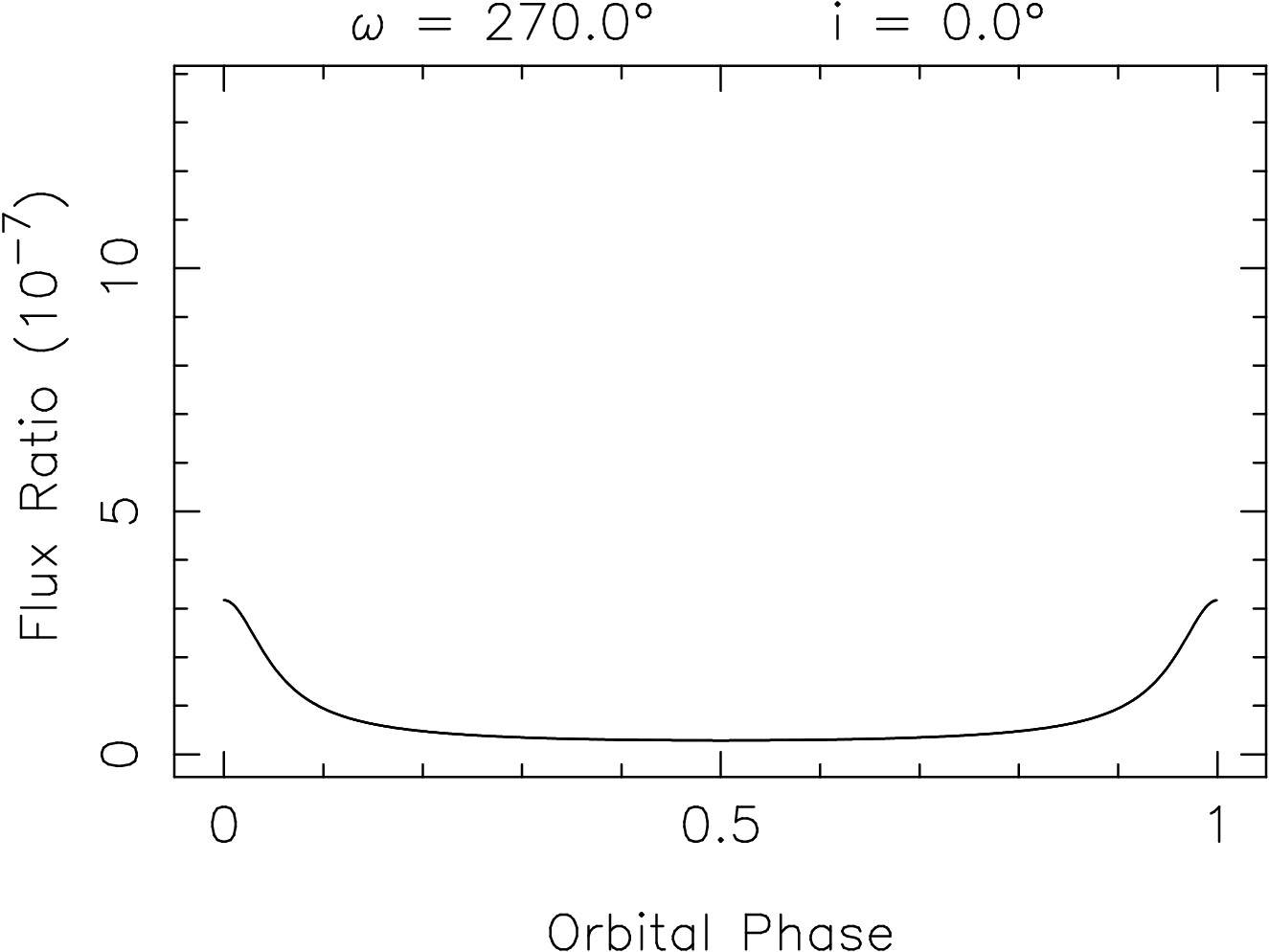} \\
    \end{tabular}
  \end{center}
  \caption{The calculated flux ratios for various orbital
    configurations with $e = 0.6$ and $P = 100$~days. Each row
    represents a specific periastron argument with inclinations
    starting from edge-on ($i = 90\degr$) and progressing to face-on
    ($i = 0\degr$).}
  \label{phases}
\end{figure*}

Beyond the cases of edge-on and face-on orbits, the interaction of
inclination and periastron argument becomes more complex. Figure
\ref{phases} demonstrates this for four eccentric orbits, each with
fixed periastron arguments, which are inclined from edge-on to
face-on. Within the range of $45\degr \lesssim \omega \lesssim
135\degr$ the flux ratio actually increases with decreasing orbital
inclination as the dayside of the planet becomes more visible where
the planet is the hottest, most noticable for the case of $\omega =
90\degr$. Outside of this regime, the peak flux from the planet
generally declines as the inclination increases and thus the access to
the full phase diminishes.

\subsection{Peak Flux Ratio Maps}

A further level of detail to the description of inclination dependence
may be added by calculating the peak flux ratio for the full range of
inclinations ($0\degr < i < 90\degr$) and periastron arguments
($0\degr < \omega < 360\degr$) for a given period and
eccentricity. The resulting intensity maps then show the optimal
orbital configuration for detection and how the distribution of peak
flux ratios smoothly varies with these configurations.

\begin{figure*}
  \begin{center}
    \begin{tabular}{cc}
      \includegraphics[angle=270,width=8.2cm]{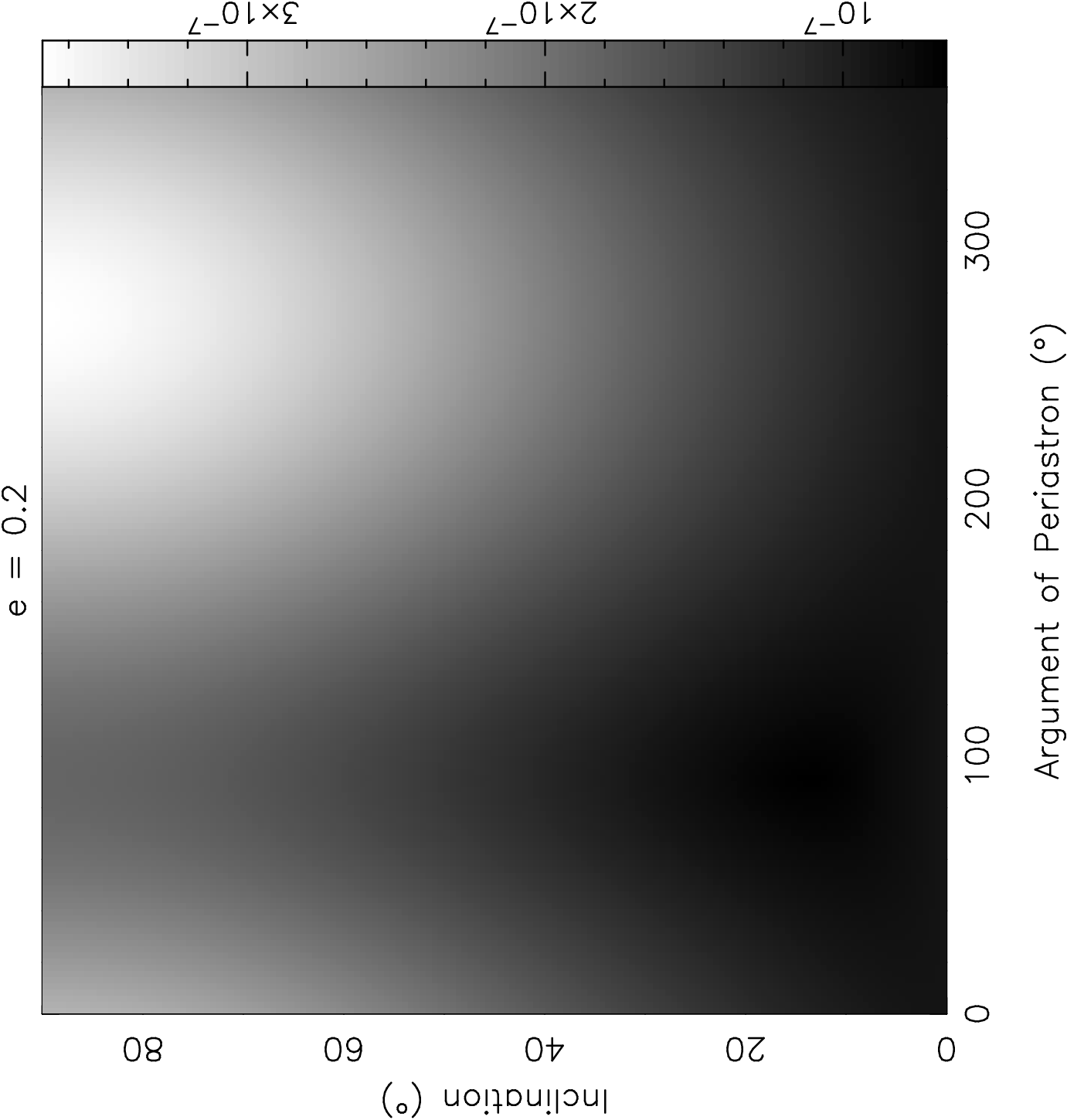} &
      \includegraphics[angle=270,width=8.2cm]{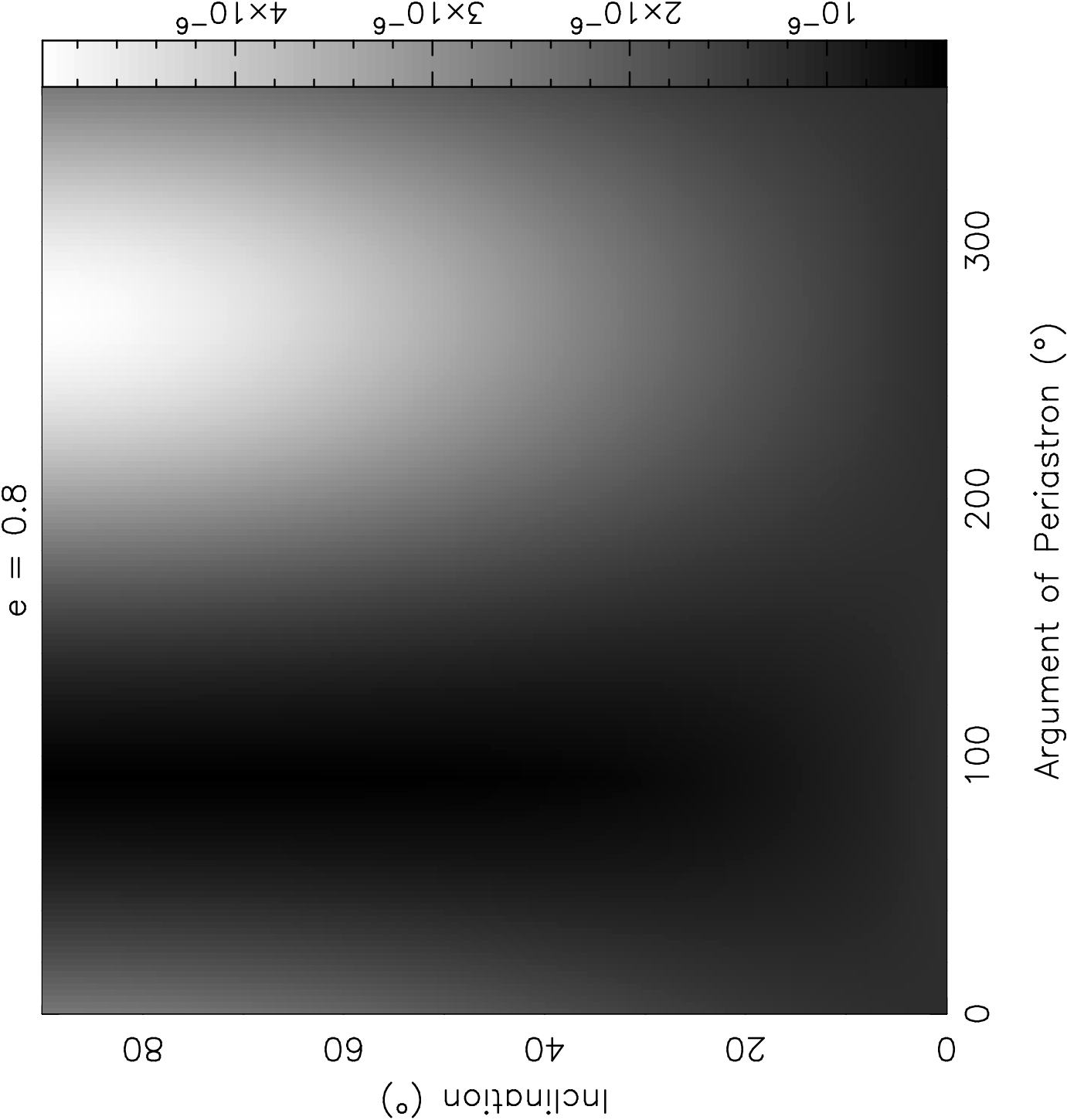} \\
    \end{tabular}
  \end{center}
  \caption{Intensity maps for a $P = 100$~days planet with $e = 0.2$
    (left) and $e = 0.8$ (right). These maps show the peak flux ratio
    values for the full range of inclinations and periastron
    arguments. For $e = 0.2$, the minimum peak flux ratio occurs where
    $\omega = 90\degr$ and $i = 13\degr$.}
  \label{array}
\end{figure*}

Shown in Figure \ref{array} are two such examples of these intensity
maps for eccentricities of 0.2 and 0.8. In each case, the strongest
flux ratio occurs where the full phase of the planet coincides with
the smallest star--planet separation ($i = 90\degr$, $\omega =
270\degr$). However, notice the interaction which occurs where $\omega
= 90\degr$. The changing peak planetary flux along this line of the
intensity maps is due to the competing components of the phase
function and star--planet separation as the planet moves from a
crescent phase near periastron to the full phase of the planet at
apastron. For highly eccentric orbits, the planet flux at $\omega =
90\degr$ becomes the brightest when the orbit is viewed face-on since
the flux ratio is dominated by the star--planet separation.

\section{Projected Separation}
\label{prosep}

The star--planet projected separation is a component which will
influence the target selection for coronagraph experiments for direct
detection of the reflected planetary flux. \citet{bei10} tabulate the
inner working angle of selected future ground-based imaging
instruments which range from $0.03\arcsec$ to $0.17\arcsec$, compared
to $0.035\arcsec$ to $0.850\arcsec$ for James Webb Space Telescope
(JWST) instruments. The angular projected separation of the planet
from the star is given by
\begin{equation}
  \Delta \theta = \frac{r}{d} \left( \cos^2 (\omega + f) + \sin^2
  (\omega + f) \cos^2 i \right)^{\frac{1}{2}}
\end{equation}
where $d$ is the star--observer distance. Here we utilize the analysis
of the previous section to determine the orbital location of maximum
flux ratio for a subset of the known exoplanets. In Table
\ref{phasetab}, we report the true anomaly $f$ (angle between the
direction of periapsis and the current position of the planet in the
orbit), flux ratio $\epsilon$, and projected separation $\Delta
\theta$ at this location for the most eccentric exoplanets. We
additionally report if the maximum flux ratio occurs for an edge-on or
face-on orientation. Notice that for face-on orbits the true anomaly
is zero at this location since the flux is completely driven by the
event of periastron passage.

\begin{table}
  \begin{center}
    \caption{$f$ and $\Delta \theta$ at maximum $\epsilon$.}
    \label{phasetab}
    \begin{tabular}{@{}lcccccc}
      \hline Planet & $P$ (d) & $e$ & $\omega$ ($\degr$) & $f$ ($\degr$) & $\epsilon (10^{-5})$ & $\Delta \theta$ (\arcsec)  \\
      \hline
HD 80606 b  &    111.4 &  0.93 &  300.6 & $ 339.7^{eo}$ & 3.80295 & 0.00003 \\
HD 20782 b  &    585.9 &  0.93 &  147.0 & $  65.4^{eo}$ & 0.09461 & 0.00336 \\
HD 4113 b   &    526.6 &  0.90 &  317.7 & $ 330.1^{eo}$ & 0.19394 & 0.00091 \\
HD 156846 b &    359.5 &  0.85 &   52.2 & $   0.0^{fo}$ & 0.03101 & 0.00359 \\
HD 45350 b  &    963.6 &  0.78 &  343.4 & $ 311.8^{eo}$ & 0.01603 & 0.00441 \\
HD 30562 b  &   1157.0 &  0.76 &   81.0 & $   0.0^{fo}$ & 0.00380 & 0.02126 \\
HD 20868 b  &    380.9 &  0.75 &  356.2 & $ 306.1^{eo}$ & 0.03875 & 0.00332 \\
HD 37605 b  &     54.2 &  0.74 &  211.6 & $  36.8^{eo}$ & 0.56082 & 0.00063 \\
HD 222582 b &    572.4 &  0.73 &  319.0 & $ 326.9^{eo}$ & 0.02634 & 0.00258 \\
HD 2039 b   &   1120.0 &  0.71 &  344.1 & $ 308.3^{eo}$ & 0.00912 & 0.00277 \\
iota Dra b  &    511.1 &  0.71 &   91.6 & $   0.0^{fo}$ & 0.00771 & 0.01182 \\
HD 96167 b  &    498.9 &  0.71 &  285.0 & $ 348.6^{eo}$ & 0.02805 & 0.00029 \\
HD 86264 b  &   1475.0 &  0.70 &  306.0 & $ 333.1^{eo}$ & 0.00779 & 0.00206 \\
HAT-P-13 c  &    428.5 &  0.69 &  176.7 & $  61.3^{eo}$ & 0.01674 & 0.00123 \\
HD 159868 b &    986.0 &  0.69 &   97.0 & $   0.0^{fo}$ & 0.00328 & 0.01054 \\
HD 17156 b  &     21.2 &  0.68 &  121.9 & $   0.0^{fo}$ & 0.33167 & 0.00068 \\
16 Cyg B b  &    798.5 &  0.68 &   85.8 & $   0.0^{fo}$ & 0.00417 & 0.02497 \\
HD 89744 b  &    256.8 &  0.67 &  195.1 & $  49.5^{eo}$ & 0.02977 & 0.00380 \\
HD 39091 b  &   2151.0 &  0.64 &  330.2 & $ 315.4^{eo}$ & 0.00429 & 0.02000 \\
HD 131664 b &   1951.0 &  0.64 &  149.7 & $  81.7^{eo}$ & 0.00236 & 0.01908 \\
HD 74156 b  &     51.6 &  0.63 &  176.5 & $  59.6^{eo}$ & 0.15667 & 0.00116 \\
HD 154672 b &    163.9 &  0.61 &  265.0 & $   4.7^{eo}$ & 0.07147 & 0.00002 \\
HD 171028 b &    538.0 &  0.61 &  305.0 & $ 334.2^{eo}$ & 0.01717 & 0.00076 \\
HD 16175 b  &    990.0 &  0.60 &  222.0 & $  36.4^{eo}$ & 0.00744 & 0.00317 \\
HD 3651 b   &     62.2 &  0.60 &  245.5 & $  17.5^{eo}$ & 0.24309 & 0.00133 \\
HD 175167 b &   1290.0 &  0.54 &  342.0 & $ 305.2^{eo}$ & 0.00450 & 0.00570 \\
HIP 2247 b  &    655.6 &  0.54 &  112.2 & $ 119.9^{eo}$ & 0.00354 & 0.02233 \\
HD 190228 b &   1136.1 &  0.53 &  101.2 & $   0.0^{fo}$ & 0.00127 & 0.01983 \\
HD 108147 b &     10.9 &  0.53 &  308.0 & $ 333.3^{eo}$ & 1.36640 & 0.00026 \\
CoRoT-10 b  &     13.2 &  0.53 &  218.9 & $  35.3^{eo}$ & 0.01872 & 0.00004 \\
HD 87883 b  &   2754.0 &  0.53 &  291.0 & $ 343.2^{eo}$ & 0.00328 & 0.00682 \\
HD 142022 b &   1928.0 &  0.53 &  170.0 & $  72.2^{eo}$ & 0.00252 & 0.02467 \\
HD 168443 b &     58.1 &  0.53 &  172.9 & $  65.4^{eo}$ & 0.09881 & 0.00244 \\
HD 81040 b  &   1001.7 &  0.53 &   81.3 & $   0.0^{fo}$ & 0.00186 & 0.02773 \\
HIP 5158 b  &    345.7 &  0.52 &  252.0 & $  14.2^{eo}$ & 0.02411 & 0.00070 \\
HD 4203 b   &    431.9 &  0.52 &  329.1 & $ 315.8^{eo}$ & 0.01272 & 0.00207 \\
HD 217107 c &   4270.0 &  0.52 &  198.6 & $  49.2^{eo}$ & 0.00117 & 0.05555 \\
HAT-P-2 b   &      5.6 &  0.52 &  185.2 & $  57.0^{eo}$ & 2.55946 & 0.00016 \\
HD 1237 b   &    133.7 &  0.51 &  290.7 & $ 344.2^{eo}$ & 0.06497 & 0.00120 \\
HD 142415 b &    386.3 &  0.50 &  255.0 & $  12.3^{eo}$ & 0.01697 & 0.00074 \\
      \hline
    \end{tabular}
    \tablecomments{$^{eo}$ indicates edge-on orbit, $^{fo}$
      indicates face-on orbit.}
  \end{center}
\end{table}

\begin{figure}
  \includegraphics[angle=270,width=8.2cm]{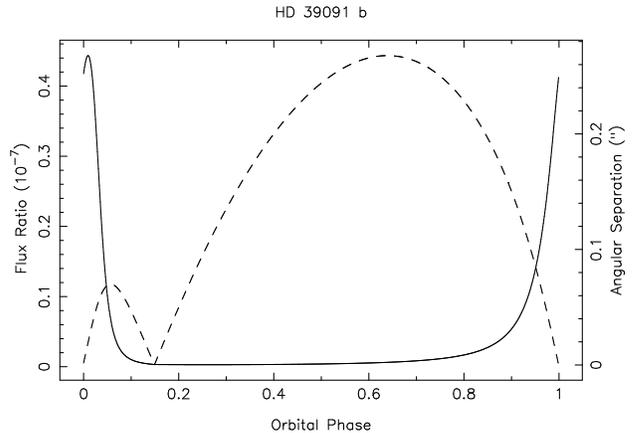}
  \caption{The flux-ratio (solid line) and star--planet angular
    separation (dashed line) for HD~39091b, assuming an edge-on
    orbit.}
  \label{fluxsep}
\end{figure}

There is a bias against optimal separation since maximum flux
naturally occurs when the planet is closest to the star. This is most
severe for edge-on orbits, such as HD~80606b where the angular
separation is $\sim 0\arcsec$ at this location. For this reason, the
location of maximum flux ratio preferentially corresponds to the
location of minimum angular separation. Consider the case of HD~39091b
(Figure \ref{fluxsep}) where the peak flux arises from an edge-on
orientation. The peak flux ratio corresponds to a $0.02\arcsec$
separation, whereas the planet reaches a separation of $\sim
0.26\arcsec$ during the entire phase. By matching the angular
separations above the resolution criteria of the coronagraph, one can
choose the optimal targets for monitoring during predicted the orbital
phase of peak flux ratio.

\section{Exoplanet Characterization}

The main targets for which phase signature detection will be attempted
will likely be those for which a planet is already known to be
present. In this case, the purpose of the observations are
chatacterization rather than discovery. As shown by Equation
\ref{fluxratio}, the flux ratio is a function of the geometric albedo
$A_g$, the periastron argument $\omega$, the orbital inclination $i$,
and the planetary radius $R_p$. The challenge of disentangling those
components from the phase shape and amplitude will vary depending upon
what is already known for that system.

If the planet is known to transit then one can measure $i$, $\omega$,
and $R_p$. One can thus determine $A_g$ which, along with models of
the planetary structure based upon the size and mass, can constrain
the properties of the atmosphere. If however the planet does not
transit, then we can estimate the radius of the planet from the
measured mass using theoretical models, such as those of \citet{bod03}
and \citet{for07}, for which there will be associated uncertainties
depending upon factors such as the age of the planet and the assumed
core model. In some cases, astrometric measurements of the host star
can performed, such as those carried out by \citet{han01}. This will
have the simultaneous result of resolving the inclination of the
orbit, thus breaking the degeneracy of the flux ratio with the
periastron argument, and determining the true mass of the secondary
companion. For non-transiting planets discovered using the radial
velocity method, one has the advantage of a strong bias towards bright
host stars relative to those discovered using the transit method. The
increase in signal-to-noise will be a great asset for an appropriate
instrument to exploit, as described in Section \ref{feasibility}.

Possible contamination of the phase signature at photometric passbands
by internal heating of the planet is worth considering. If indeed the
thermal signature of the planet is dominated by internal heating
rather than incident flux from the host star, the signature will
maintain a constant offset of the predicted photometric signature. The
self-luminous properties of the planet will be a strong function of
the age of the system, and will be restricted to planets whose age is
($< 1$~Gyr). Current radial velocity surveys choose targets based upon
stability as well as spectral line features and thus prefer F--G--K
main sequence stars where activity is known to be reduced
\citep{wri05}. Thus, targets drawn from the radial velocity
discoveries will preferentially be of an age whereby thermal
contamination of the optical phase signature will be minimal.


\section{Feasibility Requirements}
\label{feasibility}

In this section we briefly outline requirements which will influence
the detection of the previously described phase signatures.


\subsection{Instrumentation}

The detection of the phase amplitudes described here present a
significant challenge to instrumentation requirements. As mentioned in
Section \ref{introduction}, optical phase variations have been
detected in Kepler photometry. However, the vast majority of Kepler
targets are relatively faint and so not ideally suited towards
characterization with current follow-up capabilities. Conversely,
planets discovered through radial velocities have bright host stars by
comparison.

The instrumentation requirement for successful detection of the
signatures shown in Table \ref{phasetab} is photometry with an
accuracy of $\sim 10^{-6}$. The necessary stability of this precision
over long timescales depends upon the fraction of the orbital phase
over which the largest change in phase amplitude occurs, described in
detail by KG10. The photometer for the Kepler mission, for example, is
designed to achieve high-precision photometry over the 6.5 hour window
of a transit, but is not designed for long-term stability over the
lifetime of the mission \citep{bor10}. Fortunately the orbits of the
radial velocity planets are well understood in most cases and so we
can accurately predict not just the amplitude of the predicted phase
signature but also the phase and times of maximum and minimum flux
ratios. This knowledge will help to distinguish the phase signatures
from instrumental drift effects.

From the ground, the challanges are more substantial since one needs
to also contend with the offsets from night-to-night
variations. Future generation telescopes will provide opportunities to
achieve very high precision, such as the European Extremely Large
Telescope (E-ELT), the Thirty Meter Telescope (TMT), and the Giant
Magellan Telescope (GMT). It has been demonstrated by \citet{col10}
that precision photometry of $< 0.05$\% can be achieved with large
telescopes through the use of narrow-band filters.

The phase variation due to the planetary orbit is the main observable
we have discussed here but these calculations can be combined with
projected separations for use with coronagraphs, as described in
Section \ref{prosep}. Ground-based coronagraph experiments, such as
that being pursued at Palomar \citep{hin10}, are rapidly improving in
instrument efficiency and the suppression of speckle
noise. Simulations of these coronagraphs indicate that long-term
stability may indeed be possible \citep{bei10}, though this may only
be feasible in the short-term at the needed precision for young
self-luminous planets. Space-based instruments, such as the planned
Lyot coronagraph on NIRCam for the James Webb Space Telescope (JWST),
may be able to achieve phase detections for a sample of the most
favorable targets, though in this case the instrument is optimized
towards young planets around late-type stars.


\subsection{Stellar Variability}

At the level of photometric precision required here, it is important
to consider the level of intrinsic stellar variability. An analysis of
Kepler data by \citet{cia10} found that most dwarf stars are stable
down to the the precision of the Kepler spacecraft, with G-dwarfs
being the most stable of the studied spectral types. The well-known
orbital parameters of the radial velocity target stars will aid in
separating the signals of planetary phase from that of the host star
variability.  Additionally, most of the known exoplanet host stars
have been well-characterized through extensive ground-based photometry
and spectroscopy.

The main cause of photometric variability in F--G--K stars is
starspots and rotation, as verified by the Kepler variability study
performed by \citet{bas11}. The effects of starspots on exoplanet
detection have been previously studied, such as the work of
\citet{mak09}. Since the distribution of the starspot periodicity is
related to the rotation rates of these stars, the detection of
short-period planets (such as most of the known transiting planets) is
relatively unaffected by these variations. However, there will
inevitably be cases where the orbital period of the planet is close to
the rotation period of the star, which is generally in the range
10--40~days for radial velocity host stars \citep{sim10}. In such
cases, the peaks in the power spectrum from a fourier analysis of the
photometry may separate to a degree where the starspot variability can
be isolated from the phase signature. It should be noted that
disentangling these signals may substantially increase the required
observing time.


\section{Conclusions}

The push towards characterizing the atmospheres of exoplanets will
rapidly expand once further direct measurements of atmospheric albedos
and thermal properties become possible. Current space telescopes are
already detecting exoplanet phase variations in the optical (eg.,
Kepler) and the IR (eg., Spitzer). One approach to optimizing searches
for phase variations in known eccentric systems is the refinement of
orbital parameters through radial velocity measurements
\citep{kan09}. The peak flux ratio maps may then be used to constrain
the detectability of each system and the location of the peak
intensity with respect to the projected separation, particularly
important for proposed coronagraph missions such as the
Spectro-Polarimetric Imaging and Characterization of Exo-planetary
Systems (SPICES) mission. More imminently, Kepler will soon detect
transiting long-period ($P > 100$ days) planets where the bias will be
towards eccentric orbits since those have a higher probability of
transiting \citep{kan08}. In this case the results of the presented
work may be utilized to predict flux ratios for edge-on orbits as an
independent measurement of the periastron argument. The faintness of
these host stars will provide many instrumentation and observing
challenges which make known radial velocity planets more attractive
targets despite their unknown orbital inclination. Even so, as more
science results are released by the Kepler mission, the study of
photometric phase variations of long-period planets will become an
increasingly relevant endeavour in the characterization of exoplanets.


\section*{Acknowledgements}

The authors would like to thank David Ciardi and Warren Skidmore for
several useful discussions. We would also like to thank the anonymous
referee, whose comments greatly improved the quality of the paper.
This research has made use of the Exoplanet Orbit Database and the
Exoplanet Data Explorer at exoplanets.org.


\end{document}